\documentclass[preprint]{aastex}
\usepackage{epsfig,emulateapj5}

\newcommand{\km}{${\rm km\,s}^{-1}$}
\newcommand{\fuse}{{\em FUSE}}

\slugcomment{Version of \today}
\shortauthors{Lehner et al.}
\shorttitle{Deuterium toward the white dwarf WD\,0621$-$376}
\begin{document}

\title{Deuterium toward the  white dwarf WD\,0621$-$376: 
Results from the {\em Far Ultraviolet Spectroscopic Explorer} (\fuse) Mission}

\author{N.\ Lehner,\altaffilmark{1} 
	C.\ Gry,\altaffilmark{2,3}
        K.\ R.\ Sembach,\altaffilmark{1}	
	G.\ H\'ebrard,\altaffilmark{4}	
	P.\ Chayer,\altaffilmark{1,5}	
	H.\ W.\ Moos,\altaffilmark{1}
	J.\ C.\ Howk,\altaffilmark{1}	and
	J.-M.\ D\'esert\altaffilmark{4} 
	}

\altaffiltext{1}{Department of Physics and Astronomy, The Johns Hopkins University,
Bloomberg Center, 3400 N. Charles Street, Baltimore, MD 21218. nl@pha.jhu.edu}
\altaffiltext{2}{{\em ISO} Data Center, ESA, Astrophysics Division, P.O. Box 50727,
28080 Madrid, Spain.}
\altaffiltext{3}{Laboratoire d'Astronomie Spatiale, B.P. 8, F-13376 Marseille, France.}
\altaffiltext{4}{Institut d'Astrophysique de Paris, $98^{\rm bis}$ boulevard Arago, F-75014
Paris, France.}
\altaffiltext{5}{Department of Physics and Astronomy, University of Victoria, P.O. Box 3055,
Victoria, BC V8W 3P6, Canada.}

\begin{abstract}
{\em Far Ultraviolet Spectroscopic Explorer} observations are
presented for WD\,0621$-$376, a DA white dwarf star in the local
interstellar medium (LISM) at a distance of about 78 pc.
The data have a signal-to-noise ratio
of  $\sim$20--40 per 20 \km\ resolution element and cover
the wavelength range 905--1187 \AA. LISM absorption is detected in the
lines of \ion{D}{1}, \ion{C}{2}, \ion{C}{2*}, \ion{C}{3}, 
\ion{N}{1}, \ion{N}{2}, \ion{N}{3}, \ion{O}{1}, \ion{Ar}{1}, and \ion{Fe}{2}.
This sight line is partially ionized, with an ionized nitrogen fraction of  $> 0.23$. 
We determine the ratio ${\rm D/O} = (3.9 \pm\, ^{1.3}_{1.0})\times 10^{-2}$
(2$\sigma$). Assuming a standard interstellar oxygen abundance, we derive
${\rm D/H} \approx 1. 3 \times 10^{-5}$. Using the value of $N($\ion{H}{1}) derived from {\em EUVE} data
gives a similar D/H ratio.
The \ion{D}{1}/\ion{N}{1} ratio is $(3.3 \pm \, ^{1.0}_{0.8})\times 10^{-1}$ (2$\sigma$).
\end{abstract}


\section{Introduction}

Deuterium measurements place fundamental constraints on Big Bang nucleosynthesis (BBN),
the amount of baryons in the Universe, and the chemical evolution of galaxies
\citep[see][]{tytler00}. Deuterium is 
thought to be formed by primordial
BBN, and has subsequently been processed in stellar interiors. 
Therefore, the abundance of D should decrease with time and local D/H measurements 
should provide lower limits to 
the primordial D abundance. By comparing local measurements with measurements 
obtained for distant intergalactic clouds toward quasars, a better picture of the chemical
evolution of the galaxies can be drawn. A key step is to establish a robust
set of measurements for environments within the Galaxy. This has not yet been done,
and the picture that now exists is complicated by several recent studies. 
\citet{linsky98} obtained a nearly constant ratio of 
${\rm D/H} = (1.5 \pm 0.1)\times 10^{-5}$ in the Local Interstellar Cloud (LIC), while
other investigations \citep{laurent79,york83,vidal98,hebrard99,jenkins99,sonneborn00} suggest
variations amongst the different sight lines probing gas beyond the LIC. Possible causes of D/H variations 
are numerous \citep{lemoine99}, but none have yet been definitely implicated 
in causing the observed variations.

The {\em Far Ultraviolet Spectroscopic Explorer} (\fuse) is capable of 
directly observing deuterium absorption in the Lyman series lines (except Lyman-$\alpha$)
at ultraviolet (UV) wavelengths shortward of 1200 \AA. Access to higher order Lyman lines allows a larger
variety of environments (more distant, denser clouds) to be observed than is 
possible with the {\em Hubble Space Telescope} or IMAPS. 
Therefore, \fuse\/ measurements are essential for
a better description of the possible D/H variations that
may occur within the interstellar medium (ISM) of the Galaxy.

We present the first measurement of deuterium absorption toward the white 
dwarf WD\,0621$-$376 ($l,b=245.41\degr,-21.43\degr$; $V = 12.0$). The star is 
a hot ($T_{\rm eff} \approx 62,000$ K), metal-rich DA white dwarf \citep{holberg93} with a smooth stellar continuum. It 
lies at a photometric distance of $\simeq 78$ pc \citep{holberg98}. In agreement with the 
{\em IUE}\/ observations, no \ion{He}{2} absorption (e.g., $\lambda\lambda$992.3, 1084.9)
is seen in the \fuse\/ spectrum; therefore continuum placement for the 
observed \ion{D}{1} lines in the \fuse\/ band is not affected adversely
by the far-UV \ion{He}{2} series.

This article is part of a series dedicated to the first \fuse\/ measurements 
of deuterium along sight lines toward Feige\,110 \citep{friedman01}, WD\,2211$-$495 \citep{hebrard01}, 
HZ43A \citep{kruk01}, G191-B2B \citep{lemoine01}, 
BD\,+28\degr\,4211 \citep{sonneborn01}, and WD\,1634$-$573 \citep{wood01}. \citet{moos01} present 
a synthesis of these results. 

\section{\fuse\/ observations and data processing}
\fuse\/ is a dedicated observatory providing access to the 
905--1187 \AA\ wavelength range at high spectral resolution.
The design and performance of the {\fuse} spectrograph have been
described by {\citet{moos00}} and {\citet{sahnow00}}, respectively.
Two sets of \fuse\/ observations of WD\,0621$-$376 were obtained in histogram accumulation mode:
On December 6, 2000, 19 exposures totaling an exposure time of 8.3 ks
were obtained through the large $30\arcsec \times 30\arcsec$ aperture (LWRS), while 
on February 3, 2001, 21 exposures totaling 9.8 ks were obtained with the medium 
$4\arcsec \times 20\arcsec$ aperture (MDRS). The data are archived 
in the Multi-Mission Archive at the Space Telescope Institute
under the observation identifications P1041501 and P1041502.
To reduce the effects of detector fixed-pattern noise,
the data were acquired using focal plane split motions, wherein subsequent
exposures are placed at different locations on the detector.
Unfortunately, at the time of the 
observations this mode was not fully operational and 
caused misalignments of different channels during some exposures. For the LWRS observations,
8 exposures were lost in the SiC\,2 channels. For the MDRS observations, 
only the LiF\,1 channels were properly aligned; for LiF\,2, 9 exposures were lost
(for 7 exposures the detector was off);
for SiC\,1, 6 exposures were lost; and for SiC\,2 only two exposures were aligned.
Airglow emission lines were reduced along with the stellar spectra, but their intensities
were minimized as most of exposures were obtained during orbital night
(except for two exposures for each observation). 

Standard processing with the current version of the calibration
pipeline software ({\sc calfuse} v1.8.7) was used to extract
and calibrate the spectra. This version of the pipeline does
not yet correct for the astigmatism or flatfield, and therefore we required
any absorption features to be present in at least 
two channels to be considered real.  
The extracted spectra associated with the separate exposures of a given
observation were aligned by cross-correlating the positions of
strong interstellar lines, co-added, and rebinned to a nominal spectral resolution of
$\sim$$15,000$ ($\approx 20$ \km).
The interstellar lines were shifted to
0 \km\ in the laboratory rest frame \citep[note, however, that {\em IUE} spectra gave
a heliocentric velocity of $40.5 \pm 0.5$ \km,][]{holberg98}.  Including or removing the two daylight
exposures in each observation did not change the strengths of 
the airglow lines, and therefore all the exposures were co-added to maximize 
the signal-to-noise levels (S/N). 
To maintain an optimal spectral resolution and
information on the fixed-pattern noise,
neither the individual segments, nor the LWRS and MDRS observations were
co-added together. 

\begin{figure*}[tbp]
\epsscale{0.9}
\plotone{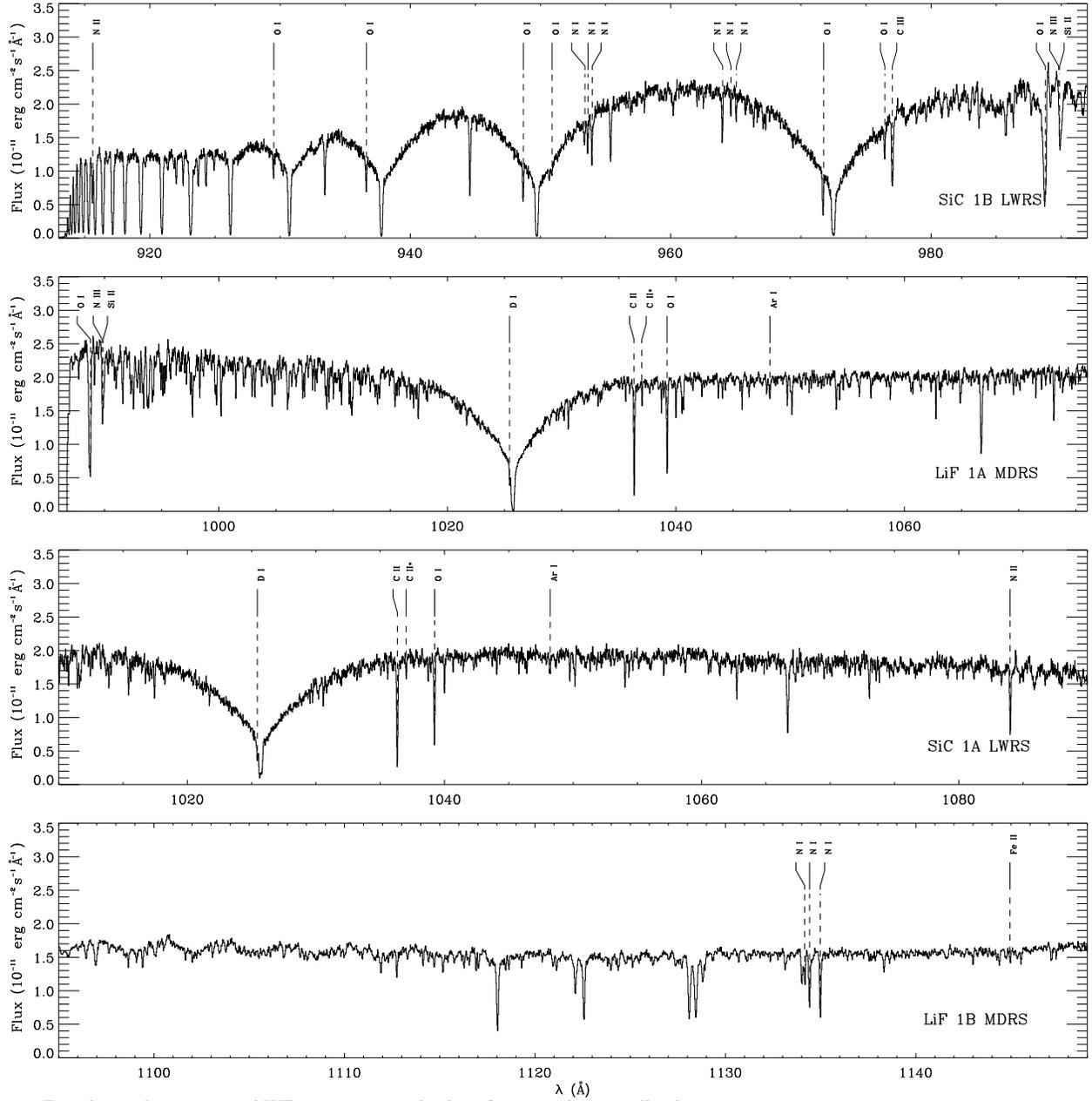}
\caption{Far-ultraviolet spectra of WD\,0621$-$376 with identification of interstellar lines. 
\label{fig1}}
\end{figure*}

We present reduced spectra for several different detector segments in 
Figure~\ref{fig1}. The data reveal numerous stellar and interstellar 
absorption lines. Prominent interstellar lines are identified.

\section{Analysis}
We used two different methods to derive the column densities of the 
interstellar species observed toward WD\,0621$-$376:
(1) The apparent optical depth method \citep[AOD,][]{savage}
was used to derive column densities and to check
for unresolved saturated structures within the observed 
profiles. A direct integration of the apparent column 
density profiles, $ N_a(v) = 3.768 \times 10^{14} \tau_a(v)/[f \lambda(\rm\AA)]$, over the 
velocity range yields the true total column 
density in cases where the absorption is weak ($\tau \lesssim 1$)
or the lines are resolved. Equivalent widths were 
directly measured by integrating the observed intensity.
(2) A profile fitting (PF) method using the code Owens
was also used to obtain the column densities of \ion{D}{1}, \ion{N}{1}, \ion{O}{1},
and \ion{Fe}{2}. The reader should refer to H\'ebrard et al. (2001) 
for a full description of the PF method, and
we will concentrate on describing in details the AOD results. The two approaches 
are different and complementary. Some of 
the issues involved in using these two methods are discussed by \citet{jenkins96}.
By using both approaches, we have a consistency check on the derived
column densities and systematic errors associated with the measurements of 
each particular method.
All methods are subject to a systematic uncertainty due to the possible 
presence of a broad component in the instrumental line spread function (LSF).  
Several tests \citep{wood01,hebrard01,kruk01} 
show however that the uncertainty due to this effect is negligible for weak metal 
lines of the type discussed in this paper.
Consequently, for the profile fitting  method, the LSF was modeled with  a 
single gaussian component. Exemples of how the LSF handling in the profile 
fitting affects the column density results are discussed in \S~\ref{pf}.

We adopted wavelengths and oscillator strengths from 
the Morton (2000, private communication) atomic data compilation. This compilation
is similar to the \citet{morton91} compilation with a few minor updates to the atomic
parameters for lines of interest in this study.

\subsection{Apparent optical depth method}\label{aod}

\subsubsection{Interstellar absorption}\label{ism}
The stellar continuum was simple enough near the \ion{C}{2}, \ion{C}{2*}, \ion{C}{3}, 
\ion{N}{1}, \ion{N}{2}, \ion{N}{3}, \ion{O}{1}, \ion{Ar}{1}, and \ion{Fe}{2} 
features to be fitted with low-order ($ \le 3$)
Legendre polynomials. Equivalent widths and apparent column densities 
are summarized in Table~\ref{t1}, where lower limits indicate 
that the line contains some unresolved, saturated absorption that cannot be reliably 
estimated with the existing data. 
Comparisons of the data from multiple channels were used to check for 
fixed-pattern noise and systematic continuum placement problems.
For the measurements the following segments were used:
SiC\,1B  (905--992 \AA); SiC\,2A (917--1005 \AA); SiC\,1A (1003--1091 \AA); 
LiF\,1A (987--1082 \AA); LiF\,1B (1094--1187 \AA); LiF\,2A (1087--1182 \AA).
The last column in Table~\ref{t1} indicates which segments were used to study each line.
In this column, the absence of a segment indicates that there may be a detector
defect at this wavelength (e.g., SiC\,2A has 
more bad pixels than SiC\,1B, especially in the range 960--965 \AA\ near the
\ion{N}{1} lines). Although the measurements were cross-checked and were
in agreement in the different segments indicated in Table~\ref{t1}, we adopted 
the values of equivalent width and apparent column density
from the segments having the best quality data.
Exceptions are the SiC\,2A LWRS and SiC\,1B LWRS and MDRS channels that have comparable quality
data; the values for lines measured on these segments were averaged 
(all the measured values were within the
estimated $1\sigma$ errors of the individual measurements).

To check if the lines contain some unresolved saturated 
structure, the apparent column density profiles of two or more lines of a given 
species with different values of $f\lambda$ were directly compared (\ion{N}{1} and \ion{O}{1}). 
The \ion{N}{1} transitions at 963.99 and 965.05
\AA\ are affected by an unknown source of noise, and this results in a 
column density that cannot be reconcilied with the values derived from other
lines (see Table~\ref{t1}). This discrepancy is unlikely to 
arise from the adopted oscillator strengths, as studies of other sight lines 
do not show any particular problem for these transitions. Since we have 
information from different channels (see last column in Table~\ref{t1}),
fixed-pattern noise also seems unlikely. Contamination
by unidentified photospheric stellar lines is the most likely source 
of confusion since this star is metal rich.
These two lines were removed from the analysis. Of the remaining \ion{N}{1} lines,
only \ion{N}{1} $\lambda$1134.98 suffers from 
weak saturation effects. Similar comparisons were 
made for \ion{O}{1}, and the transitions at 948.686, 971.738 and
1039.230 \AA\ contain unresolved saturated structures (see Table~\ref{t1}).
We note that, while the total column densities of all the unsaturated 
lines of \ion{N}{1} and  \ion{O}{1} agree remarkably well within the $1\sigma$ errors, 
the strongest \ion{N}{1} $\lambda$$\lambda$953.655, 1134.415 and \ion{O}{1} 
$\lambda$$\lambda$936.630, 976.448 lines yield the lowest apparent column densities, indicating
some weak saturation effects might be present. For the other species, assuming they 
follow a similar curve of growth, only \ion{C}{2} and \ion{N}{2} are likely to 
be influenced strongly by saturation effects.

\begin{figure}[tbp]
\epsscale{0.4}
\plotone{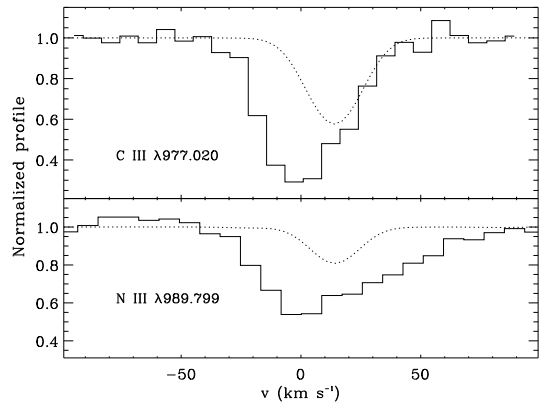}
\caption{Spectra of \ion{C}{3} and \ion{N}{3} (histograms) and a model of stellar absorption
(dotted lines). The \ion{N}{3} feature shows clearly that interstellar \ion{Si}{2} $\lambda$989.9 must also be
present.
\label{fig2}}
\end{figure}

Both \ion{C}{3} and \ion{N}{3} are blended with their respective stellar features.
Figure~\ref{fig2} shows a stellar model ($T_{\rm eff} = 62,000$ K; $\log g = 7.0$, 
$\log {\rm C/H} = -6.0$; $\log {\rm N/H} = -6.6$, preliminary results from
Chayer et al., in preparation)
that estimates the stellar contribution ($\sim$$30\%$ for \ion{C}{3}, $\sim$$7\%$ for 
\ion{N}{3}). The stellar model was aligned to the observed stellar lines, and an offset 
of $\approx +14$ \km\ was found with respect to the interstellar lines.
\ion{N}{3} suffers from blending with interstellar \ion{Si}{2} $\lambda$989.87 absorption,
which is clearly visible in Figure~\ref{fig2}, but the combination of relatively low spectral 
resolution and the proximity of the \ion{N}{3} stellar feature did not allow us to model this line.  
We examined the {\em IUE}\/ spectrum of WD\,0621$-$376 to check for \ion{Si}{2} absorption 
at longer wavelengths, but its poor quality did not shed any light on the amount of \ion{Si}{2} along  
this sight line. Therefore, the derived equivalent width of \ion{N}{3} must 
be considered an upper limit. 

The spectra also show the detection of a very weak feature \ion{Fe}{2} at 1144.938 \AA. Though
its measurement remains uncertain, its identification is unambiguous as 
the radial velocity of this feature is in agreement with the velocities of the \ion{N}{1}
lines in the LiF\,1B and LiF\,2A segments.

The $3\sigma$ upper limits for the equivalent widths in Table~\ref{t1}
are defined as $W_{\rm min} = 3 \sigma\, \delta\lambda$, where 
$\sigma$ is the inverse of continuum S/N ratio and $\delta \lambda \approx 0.1$ \AA,
which is approximately the average FWHM obtained from the resolved features.
The corresponding  $3\sigma$ upper limits on the column density
are obtained from the corresponding equivalent width limits and the assumption 
of a linear curve of growth. 

\subsubsection{Deuterium absorption}\label{deut}
\begin{figure}[tbp]
\epsscale{0.5}
\plotone{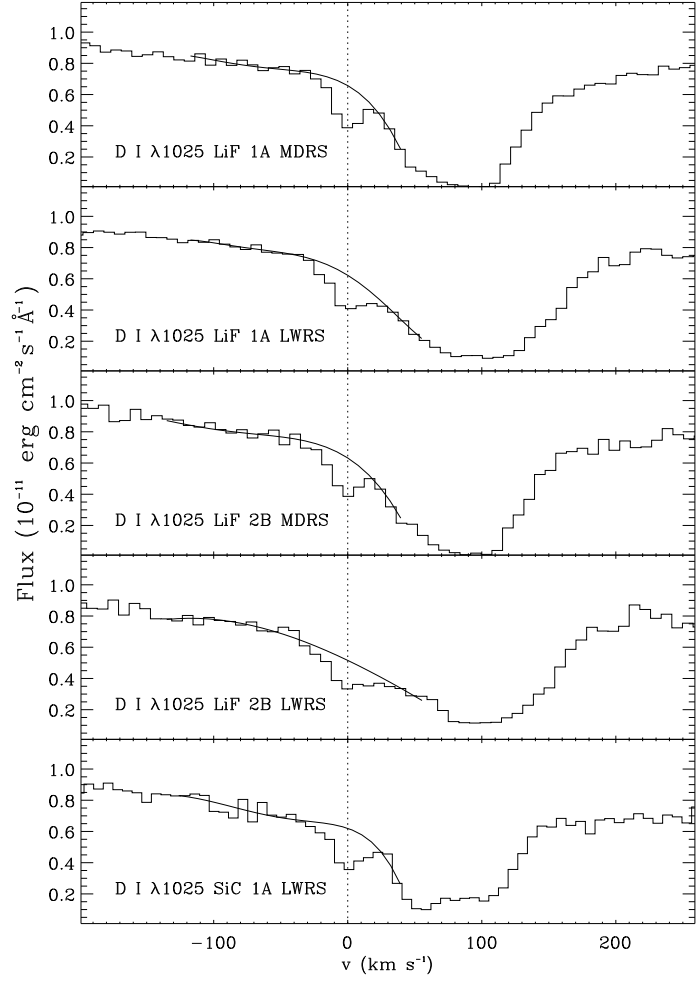}
\caption{Spectra of \ion{D}{1} Lyman-$\beta$ in the different segments and local polynomial fits to the continua.
Note how the shapes of the lines differ between segments in the LWRS data (even for essentially
night data) because of the \ion{H}{1} airglow, while for the MDRS the effect of airglow is smaller.
\label{fig3}}
\end{figure}

\begin{figure}[tbp]
\epsscale{0.5}
\plotone{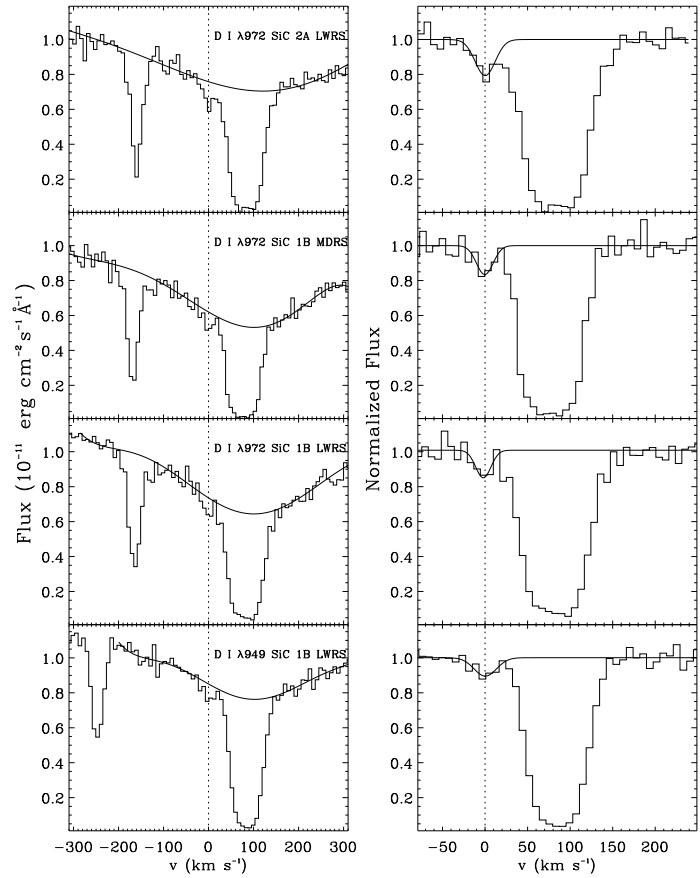}
\caption{Detection (2--3$\sigma$) of the
\ion{D}{1} lines in the SiC channels.
The left panel shows the calibrated flux in different segments (and wavelengths), while the right panel
side presents their respective normalized profile with a gaussian fit.
\label{fig4}}
\end{figure}

\ion{D}{1} Lyman-$\beta$ is well detected in the \fuse\/ spectra of WD\,0621$-$376
(see Figure~\ref{fig3}). 
Figure~\ref{fig4} shows some evidence for tentative detections of \ion{D}{1} $\lambda$972
and \ion{D}{1} $\lambda$949. Polynomial fits to the continua are shown for different segments 
along with their respective
normalized profiles with optimized gaussian fits (see below). 
Except for segment SiC\,2A, all the measurements for  \ion{D}{1} $\lambda\lambda$972 
and 949 yield detections with less than $3\sigma$ significance (see Table~\ref{t2}). For segment SiC\,2A,
both the continuum placement and the line detection remain very uncertain and 
fixed-pattern noise may contaminate this feature. The measurements of these
lines and the $3\sigma$ upper limits on $W_\lambda$ and $\log N$ were derived following the same methods
described in \S~\ref{ism}. 

Figure~\ref{fig3} displays the \ion{D}{1} Lyman-$\beta$ spectra in the 
different segments. We note that the \ion{H}{1} airglow is relatively strong and broad ($\sim$100 \km)
in the LWRS data. Its strength varies in the different segments; for these
data, the airglow in segments LiF\,2B and SiC\,1A affects the \ion{D}{1} measurement.
The MDRS observations are less affected by the airglow, but 
it is clearly present. The LiF\,2B \ion{D}{1} data are more affected by the
\ion{H}{1} airglow than the LiF\,1A data. 

The measurements for \ion{D}{1} are summarized in Table~\ref{t2}.
The continuum placement is more complicated than for the species described in \S~\ref{ism},
requiring higher order (3--5)  polynomial fits to the continuum. 
The continuum fits in the LWRS LiF\,2B and SiC\,1A segments are very
uncertain due to the airglow contamination; the measurements, however, agree 
with the other measurements (Table~\ref{t2}). Generally good agreement 
is found between the MDRS LiF\,1A, LiF\,2B and LWRS LiF\,1A data. Of these,
the  MDRS LiF\,1A is the least affected by airglow, so we adopt this result 
($\log N$(\ion{D}{1}$) = 13.75 \pm 0.09$ dex, 1$\sigma$) for comparison with the 
other value derived below.

\begin{figure}[tbp]
\epsscale{0.5}
\plotone{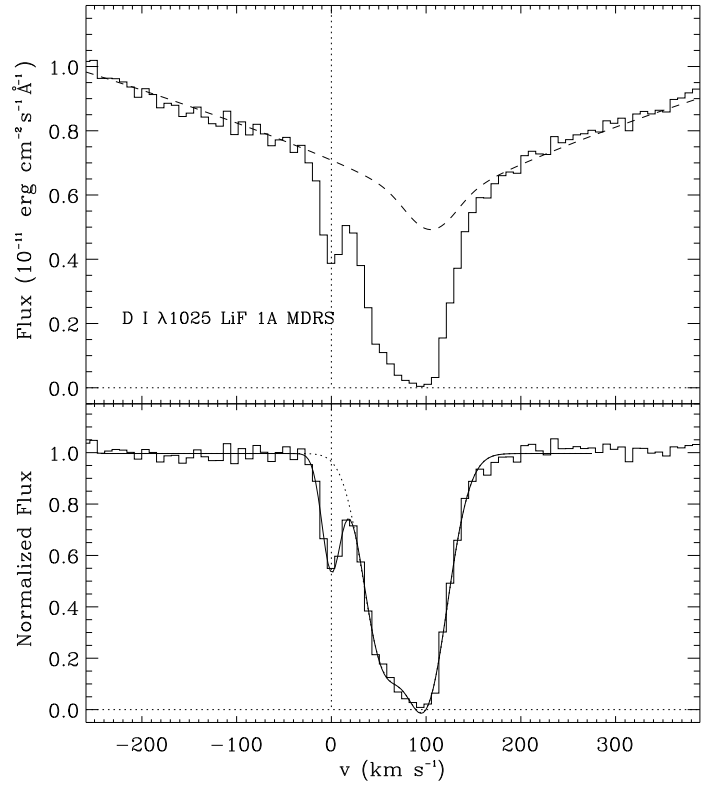}
\caption{Spectrum of \ion{D}{1} Lyman-$\beta$ in the LiF\,1A MDRS segment. Top panel: 
calibrated flux with the stellar continuum. Bottom panel: 
the resulting normalized profile with optimized gaussian fits 
(the solid line is the 3 gaussian component fit (\ion{D}{1}
and \ion{H}{1}), while the dotted line is the 2 gaussian component fit (only \ion{H}{1}); 
see text for more details). 
\label{fig5}}
\end{figure}

The continuum was also determined using the stellar model described in \S~\ref{ism}
for the MDRS LiF\,1A observations: The top panel of Figure~\ref{fig5} shows the model
overplotted on the data. While Figure~\ref{fig5} shows
a reasonable stellar model, there are still some problems
as some portions of the spectrum are not so well reproduced by the stellar
model. To estimate the contribution 
from the  \ion{H}{1} feature, the normalized profile (bottom panel of Figure~\ref{fig5}) was fitted
with 3 gaussian absorption components using an optimized routine \citep{howarth}, where the FWHM and
the centroids were free to vary; changing (in a reasonable manner) the initial guess of these parameters 
did not change the fit.  This method
does not attempt to produce any realistic physical quantities for \ion{H}{1}, but just
tests the effective continuum placement for \ion{D}{1}, as optimized gaussian fits can usually reproduce
complicated profiles \citep{lehner99}. The bottom panel of Figure~\ref{fig5} also shows 
the same fit with 2 gaussian components to the \ion{H}{1} feature (FWHMs and centroids were fixed to the 
values derived with the 3 components fit) to indicate the degree of contamination from \ion{H}{1}
to \ion{D}{1}.
This approach yields a logarithmic apparent column density of $13.84 \pm 0.10$ dex (1$\sigma$). 
Considering other segments gives similar results, but these are more uncertain 
as the airglow is stronger. 

The above estimates of $N($\ion{D}{1}) agree within their 1$\sigma$ errors, though
the difference in the absolute values shows some uncertainties caused by the continuum placement.
We adopt $\log N$(\ion{D}{1}$) = 13.79 \pm 0.14$ dex (2$\sigma$), obtained
by doing a weighted mean of the apparent column densities derived using the two 
different continuum placements. Table~\ref{t3} summarizes the definitive results
for the different ions (for \ion{N}{1} and \ion{O}{1}, the values are weighted means of 
the values listed in Table~\ref{t1})

\begin{figure}[tbp]
\epsscale{0.5}
\plotone{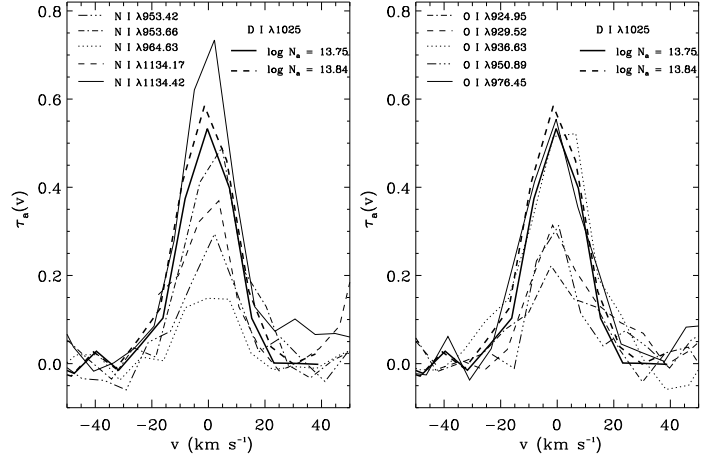}
\caption{Apparent optical depths of \ion{D}{1} Lyman-$\beta$ for the two continuum placements
(see \S~\ref{deut}) compared to the unsaturated \ion{N}{1} and \ion{O}{1}.
\label{fig6}}
\end{figure}

As discussed above, the S/N levels for the higher \ion{D}{1} Lyman
series were not high enough to make precise estimates of $N($\ion{D}{1}). However, their $3\sigma$ upper
limits do not suggest that the \ion{D}{1} Lyman-$\beta$ line is strongly saturated,
if at all. Comparing the apparent optical depths of \ion{D}{1} Lyman-$\beta$  
with the unsaturated \ion{N}{1} and  \ion{O}{1} lines (see Figure~\ref{fig6})
indicates that the \ion{D}{1} Lyman-$\beta$ optical depths for the two continuum placements are only 
as strong as the strongest \ion{N}{1} and  \ion{O}{1} 
features, suggesting that the line might not be entirely resolved (see \S~\ref{ism}).
However, saturation corrections should be small ($\lesssim 0.05$ dex according to the \ion{N}{1} and
\ion{O}{1} lines) and are less than the 2$\sigma$ error bars on the measurement.

\subsection{Results of the profile fitting and comparison}\label{pf}

In addition to  estimating column densities using the apparent optical depth method, two
independent profile fitting analyses using voigt profiles and using the information 
simultaneously from all the windows were performed, leading
to two sets of results, PF\,1 and PF\,2.
In PF\,1, the profile fitting code Owens was used to fit the data in 32 spectral windows, including
all available unsaturated lines from \ion{D}{1}, \ion{N}{1}, \ion{O}{1} and
\ion{Fe}{2} (i.e., \ion{N}{1} $\lambda$1134.980,  \ion{O}{1}
$\lambda$$\lambda$948.686, 971.738 and 1039.230 were excluded).
In PF\,2, a total of 24 windows
were used including lines from \ion{D}{1}, \ion{N}{1}, and \ion{O}{1}. 
For Lyman-$\beta$,
only the MDRS data were used in both cases because the LWRS data are significantly
affected by airglow. The results for these two methods are
listed in Table~\ref{t3}.
Although  there are differences both
in the absolute values and in the estimated 2$\sigma$ errors,
all results agree within 1$\sigma$.
Apart from the different number of windows used, the differences in the results
arise from the strategies adopted to deal with the
widths of the LSF. The LSFs are not yet well
characterized and  are known to vary from one segment to another. In PF\,1 the LSF widths of 
about 60\% of the
windows were fixed to a pre-defined value (ranging from 8.7 to 12 pixels,
depending on the window), while in PF\,2 the
LSF widths were free
to vary, with the constraint that they remain within plausible values
(in the accepted fits the maximum FWHM was 15 pixels and the minimum was 6 
pixels).\footnote{Note that the dispersions of the FUSE spectrographs are
$\sim$6.7 and $\sim$6.2 m\AA\ per pixel for LiF and SiC channels, respectively.}

The largest differences, found for the \ion{N}{1} column densities, are due
to the fact that in PF\,2 high column density values -- implying saturation 
of some of the fitted \ion{N}{1} lines -- were permitted by the adoption of broader 
LSFs.

\begin{figure}[tbp]
\epsscale{0.5}
\plotone{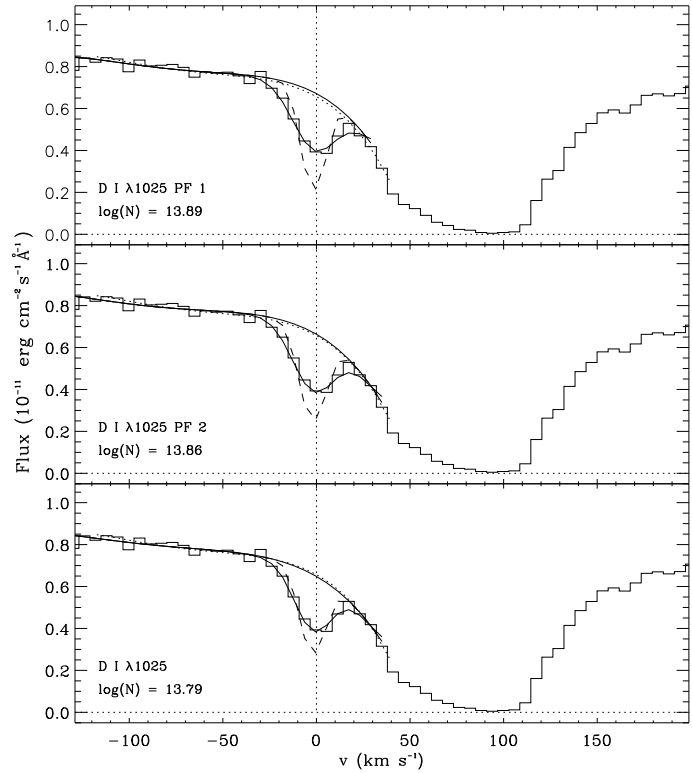}
\caption{Spectrum of \ion{D}{1} Lyman-$\beta$ in the LiF\,1A MDRS segment with fits from PF\,1 and PF\,2 (see \S~\ref{pf}).
The fitted profiles as well as the continua are drawn by a solid line
while the dashed line shows the intrinsic profiles before convolution by the LSFs.
The bottom panel shows the value derived with an Owens fit using the PF\,2 
method but with the \ion{D}{1} column density set to the 
value (13.79 dex) derived from the AOD method. The dotted lines shown in the 
three panels indicate the continuum shown in the top panel of Figure~\ref{fig3}. 
\label{fig7}}
\end{figure}

The top and middle panels of Figure~\ref{fig7} show the best fits to the \ion{D}{1} Lyman-$\beta$ 
line in the LiF\,1A MDRS channel, as obtained by PF\,1 ($\log N$(\ion{D}{1}$) =13.89$ dex) 
and PF\,2 ($\log N$(\ion{D}{1}$) =13.86$ dex), respectively.
As expected, there is no difference in the fitted profiles, but the
difference in the adopted fit parameters can be seen in the profiles plotted before convolution with the 
LSF. The  unconvolved
profile  is slightly narrower (and deeper) in PF\,1 than in PF\,2
because of a slightly lower $b$-value (7.1 vs 8.9 \km) required to compensate for a slightly higher
LSF width (10.5 vs 9.4 pixels in the LiF\,1A MDRS data).

The bottom panel of Figure~\ref{fig7} shows a fit performed with Owens using the PF\,2 method 
but fixing the \ion{D}{1} column density to the best value ($\log N$(\ion{D}{1}$) =13.79$ dex) 
derived in \S~\ref{deut} using the 
AOD method. In this case, the unconvolved profile is shallower than the other two, 
but the convolved profile is not, because this is compensated by a narrow LSF (8 pixels).
For the AOD value the optical depth of the Lyman-$\beta$ line is smaller
than 1. For the two best fits in PF\,1 and PF\,2, the Lyman-$\beta$ line is slightly
saturated, explaining why the \ion{D}{1} column density results obtained by
the two profile fitting studies are higher than those obtained by the AOD method.

The results obtained by the three independent investigations are shown
in Table~\ref{t3}. They all agree within the estimated 1$\sigma$ errors. Our
adopted result, shown in the last column of Table~\ref{t3}, is the simple mean
between these three results. 

\section{Neutral and partially ionized gas}
\ion{O}{1} is an excellent tracer of neutral gas as
its ionization potential and charge exchange reactions with hydrogen
ensure that the ionization of \ion{H}{1} and \ion{O}{1} are strongly coupled. $N($\ion{O}{1})/$N($\ion{H}{1})
is therefore an excellent approximation of the ratio O/H. (Similarly the ratio D/O is well approximated 
by $N($\ion{D}{1})/$N($\ion{O}{1})). Therefore, 
using the \citet{meyer98} interstellar O abundance\footnote{Meyer et al.'s  value 
($ {\rm O/H} = 
(3.19 \pm 0.14)\times 10^{-4}$) was corrected for
the recommended oscillator strength of \ion{O}{1} $\lambda$1356 ($f= 1.16 \times 10^{-6}$
instead of $f= 1.25 \times 10^{-6}$) by 
\citet{welty99}.}, $ {\rm O/H} = 
(3.43 \pm 0.15)\times 10^{-4}$, and the \ion{O}{1} column density
along the WD\,0621$-$376 sight line, we derive a total \ion{H}{1} column density
of 18.73 dex. 
We note, however, that the Meyer et~al. sight lines typically
sample more distant (100--1000 pc) gas than the WD\,0621$-$376 sight line.
From the \fuse\/ spectra, it 
is not possible to derive a reliable \ion{H}{1} column density,
as the \ion{H}{1} transitions are essentially confined on
the flat part of the curve of growth for this sight line. Previously, \citet{holberg98}
quoted $\log N$(\ion{H}{1})~$=18.15 \pm 0.15$ dex (1$\sigma$), which 
is not in agreement with our value derived above. Using {\em EUVE}\/
spectra, \citet{wolf98} found $18.70$ dex, in good
agreement with the value derived from \ion{O}{1}, but unfortunately
no error is quoted. Considering the {\em IUE}\/ spectrum,
we note that it is not possible to obtain \ion{H}{1} from these data due its relatively 
low quality.  Only a combination of a very high S/N spectrum and a very good
stellar model would help to directly determine the \ion{H}{1} column density to a
reasonable degree of accuracy.

With the derived \ion{N}{1} column density and the \citet{meyer97} 
interstellar N abundance, $ {\rm N/H} = (7.5 \pm 0.4)\times 10^{-5}$, 
one finds 18.47 dex for \ion{H}{1}, a value of about 48\% lower than the value
derived from \ion{O}{1}. The 
\fuse\/ bandpass allows us to directly estimate the quantity
of ionized nitrogen, as \ion{N}{2} and \ion{N}{3} are accessible. \ion{N}{3} is too poorly
determined to be of much use as the amount of \ion{Si}{2} remains unknown. 
However, the ratio \ion{N}{1}/(\ion{N}{1}+\ion{N}{2}$) < 0.77$
again suggests that a substantial fraction ($>$$23\%$) of N is ionized.
Therefore, the discrepancy in values of $N($\ion{H}{1}) derived from \ion{N}{1}
and \ion{O}{1} may be due in part to ionization of N in \ion{H}{1} gas along 
the sight line. While the charge exchange with hydrogen should make \ion{N}{1} 
a good indicator of the amount of \ion{H}{1} along the sight line 
\citep{sofia98}, \citet{jenkins00} noted that in regions where 
$n_e \gg n_{\rm H^0}$, N may show a deficiency of its neutral 
form, similar to what is observed for argon.

A similar result holds for the ionization of argon. 
Using our derived \ion{Ar}{1} column density and the Ar abundance from \citet{anders},
$ {\rm Ar/H} = (3.6 \pm\, ^{0.9}_{0.7}) \times 10^{-6} $, we derive an \ion{H}{1} 
column density of about 18.30 dex, 
which, when compared to the value derived from \ion{O}{1}, implies that 65\%
of the argon is ionized (neither Ar nor N is depleted significantly onto
dust grains). These results for both \ion{N}{1} and \ion{Ar}{1} indicate 
that photoionization may be an important process along this sight line. 
Towards four other nearby white dwarfs \cite{jenkins00}, similar results were found, favoring the idea
that the LISM ionization is maintained by a strong extreme ultraviolet flux (yet to be
identified) rather than 
an incomplete recombination of helium from a previous episode of 
collisional ionization at high temperatures. The presence of \ion{C}{3} also implies that 
the gas is partially ionized. \ion{Fe}{2} is largely deficient ($-1.3$ dex), 
but this is at least partly explained by iron being depleted onto dust grains.

\section{D/O, D/N and D/H ratios}\label{ratio}
Using the results from Table~\ref{t3}, and following the above discussion 
(D/O~=~\ion{D}{1}/\ion{O}{1}), we calculate  
${\rm D/O} = (3.9 \pm\, ^{1.3}_{1.0})\times 10^{-2}$ ($2\sigma$).
Using the interstellar O abundance, the D/H ratio can be derived: 
${\rm D/H} \approx  1.3 \times 10^{-5}$. A similar D/H ratio (${\rm D/H} \sim  1.4 \times 10^{-5}$)
is obtained using $N$(\ion{H}{1}) derived from the {\em EUVE} spectra.

The \ion{D}{1}/\ion{N}{1} ratio is $(3.3 \pm \, ^{1.0}_{0.8})\times 10^{-1}$ ($2\sigma$).
As there is a non negligible fraction of ionized nitrogen and some of this fraction can
be either in \ion{H}{1} or \ion{H}{2} regions, it is not possible
to derive the ratio D/H using the nitrogen abundance.

Toward WD\,0621$-$376, the \ion{H}{1} column density
remains relatively uncertain, but D/H is consistent with 
the LIC ratio 
\citep[$(1.5 \pm 0.1)\times 10^{-5}$,][]{linsky98} and also consistent with other 
measurements within the LISM \citep{moos01}. 
Since in the \fuse\ bandpass, \ion{D}{1} and \ion{O}{1} can be generally
measured simultaneously, D/O ratio (more precisely measured) can be used instead of D/H. To be noted
that if D/O becomes available in distant quasars, the D/O  ratio would be very
sensitive to astration since \ion{D}{1} is destroyed while \ion{O}{1}
is enhanced through the various stellar cycles. Initial 
measurements of D/O ratios in the LISM \citep{moos01} show no significant variations
which also favors constant D/H and O/H ratios in the LISM.
Other results suggest variations at larger distances \citep[e.g.,][]{jenkins99,sonneborn00}, 
but more observations beyond the LISM wall are needed to have a better understanding
of these possible variations in the ISM. 
\section{Summary}
We have presented \fuse\/ observations of the DA white-dwarf star WD\,0621$-$376, located
at about 78 pc. The spectra show interstellar absorption from  
\ion{D}{1}, \ion{C}{2}, \ion{C}{2*}, \ion{C}{3}, 
\ion{N}{1}, \ion{N}{2}, \ion{N}{3}, \ion{O}{1}, \ion{Ar}{1}, and 
\ion{Fe}{2}. Using both the N ISM abundance and the
\ion{N}{1} and \ion{N}{2} column densities, we showed that
N is ionized by a fraction $>$$23\%$ (Ar is ionized by about $65\%$). 
These results confirm the recent suggestion by
\citet{jenkins00} that a diffuse ionizing flux ($> 24.6$ eV) 
creates the high fraction of ionized helium observed in the LISM.

The D/O ratio along the WD\,0621$-$376 sight line is  $ (3.9 \pm\, ^{1.3}_{1.0})\times 10^{-2}$ ($2\sigma$),
implying a D/H ratio of approximately
$ 1.3 \times 10^{-5}$.  This D/H ratio is also
supported by the \ion{H}{1} measurement obtained from {\em EUVE}.
The \ion{D}{1}/\ion{N}{1} ratio is $(3.3 \pm \, ^{1.0}_{0.8})\times 10^{-1}$ ($2\sigma$).
Both D/O and D/H ratios are in agreement with other LISM measurements, implying that D/O 
and D/H are relatively homogeneous in the LISM

\acknowledgements
This work is based on data obtained
for the Guaranteed Time Team by the NASA-CNES-CSA FUSE mission operated 
by the Johns Hopkins University. Financial support to U. S.
participants has been provided by NASA contract NAS5-32985.
French participants are supported by CNES.
This work was partially done using the profile fitting procedure Owens.f
developed by M. Lemoine and the FUSE French Team.

\clearpage

\begin{deluxetable}{lcccccc}
\tablecolumns{7}
\tablewidth{0pc} 
\tablecaption{Interstellar absorption lines (but \ion{D}{1}): Equivalent widths and AOD results  \label{t1}} 
\tablehead{\colhead{Ions}    &   \colhead{$\lambda_{\rm lab}$$^a$} &   \colhead{$f$$^a$} &
\colhead{S/N$^b$} &\colhead{$W_\lambda$} & \colhead{$\log N_a$}& \colhead{Segment$^b$} \\ 
\colhead{}  &\colhead{\AA} & \colhead{} &\colhead{} &\colhead{(m\AA)}& \colhead{(${\rm cm}^{-2}$)}& \colhead{} }
\startdata
\ion{C}{2}  	 & 1036.337  &   $1.18 \times 10^{-1}$    &	33	& $ 98.9 \pm 2.5$	& $> 14.19 $		  &  1,(2,3)	\\
\ion{C}{2*}  	 & 1037.018  &   $1.25 \times 10^{-1}$    &	54	& $ 12.8 \pm 1.5$	& $  13.06 \pm 0.05 $	  &  1,(2,3)   	\\
\ion{C}{3}  	 & 977.020   &   $7.62 \times 10^{-1}$    &	24	& $ 62.5 :$$^c$		& $> 13.14 :  $$^c$	  &  4,5     	\\
\ion{N}{1}  	 & 953.415   &   $1.32 \times 10^{-2}$    &	32	& $ 21.5 \pm 3.0$	& $14.34 \pm 0.07 $	  &  4,6    	\\
	       	 & 953.655   &   $2.50 \times 10^{-2}$	  &     30      & $ 32.9 \pm 5.0$       & $14.29 \pm 0.07 $	  &  4,5,6   	\\
		 & 953.970   &   $3.48 \times 10^{-2}$	  &     30      & $ 41.2 : $$^d$       	& $14.24:$$^d$	  	  &  4,5,6  	 \\
	         & 963.990   &   $1.48 \times 10^{-2}$	  &     37      & $ 30.8\pm  2.3$       & $14.47 \pm 0.04 $	  &  6,(4)    	 \\
	         & 964.626   &   $9.43 \times 10^{-3}$	  &     30      & $ 16.4 \pm 2.9$       & $14.35 \pm 0.08 $	  &  5,6  	 \\
	         & 965.041   &   $4.02 \times 10^{-3}$    &  	32      & $ 16.0 \pm 3.3$       & $14.72 \pm 0.15 $	  &  5,6		\\
	         & 1134.165  &   $1.52 \times 10^{-2}$    &     48      & $ 34.1 \pm 3.1$       & $14.35 \pm 0.04 $	  &  8,(7) 	\\
	         & 1134.415  &   $2.97 \times 10^{-2}$    &     35      & $ 51.6 \pm 2.8$       & $14.31 \pm 0.04 $	  &  8,(7)  	\\
	         & 1134.980  &   $4.35 \times 10^{-2}$    &     44      & $ 56.2 \pm 2.8$       & $> 14.18  $	 	  &  8,(7)   	\\
\ion{N}{2}  	 & 915.613   &   $1.65 \times 10^{-1}$    &	20	& $ 51.5 :$		& $> 13.74 :  $	  	  &  5,6	\\
		 & 1083.994  &   $1.15 \times 10^{-1}$    &     40      & $ 58.2 \pm 2.8$       & $> 13.80  $	  	  &  3,(9)   	\\
\ion{N}{3}  	 & 989.799   &   $1.23 \times 10^{-1}$    &	21	& $ < 79.0: $$^e$	& $ < 13.94 : $$^e$	  &  1,(2,3)   	\\
\ion{O}{1}  	 & 924.950   &   $1.54 \times 10^{-3}$    &	26	& $ 17.8 \pm 3.9$	& $15.23 \pm 0.08 $	  &  5,6,(4)     \\
		 & 929.517   &   $2.29 \times 10^{-3}$    &	25	& $ 21.8 \pm 3.9$	& $15.22 \pm 0.08 $	  &  4,5,6     \\        
		 & 936.630   &   $3.65 \times 10^{-3}$    &     26      & $ 37.1 \pm 4.5$       & $15.20 \pm 0.10 $	  &  4,5,6   \\         
		 & 948.686   &   $6.31 \times 10^{-3}$	  &     23      & $ 57.1 \pm 4.4$       & $> 15.13 	 $	  &  4,5,6    \\
		 & 950.885   &   $1.58 \times 10^{-3}$	  &     21      & $ 18.8 \pm 4.3$       & $15.22 \pm 0.09 $	  &  4,(5,6)   \\	         
		 & 971.738   &   $1.16 \times 10^{-2}$	  &     23      & $ 69.6 \pm 4.7$       & $> 15.04 $	 	  &  4,5,6   \\
	         & 976.448   &   $3.31 \times 10^{-3}$	  &     23      & $ 34.2 \pm 6.8$       & $15.17 \pm 0.15 $	  &  4,5,6   \\
	         & 988.773   &   $4.65 \times 10^{-2}$	  &     25      &  \nodata$^f$	        &  \nodata$^f$		  &  4,(5,6)   \\
	         & 1039.230  &   $9.20 \times 10^{-3}$	  &     32      & $ 79.6 \pm 3.4$       & $> 15.11 $		  &  1,(2,3)   	\\
\ion{Si}{2}  	 & 989.873   &   $1.87 \times 10^{-1}$    &	26	& \nodata$^e$		& \nodata$^e$		  &  1,(2,3)   	\\
	  	 & 1020.699  &   $1.64 \times 10^{-2}$    &	41	& $ < 7.3 $		& $< 13.68 $		  &  1,(2,3)   	\\
\ion{Ar}{1}  	 & 1048.220  &   $2.63 \times 10^{-1}$    &	41	& $ 17.6 \pm 2.8$	& $12.86 \pm 0.07 $	  &  1,(2,3)   	\\
	  	 & 1066.660  &   $6.65 \times 10^{-2}$    &	40	& \nodata$^g$		& \nodata$^g$		  &  1  	\\
\ion{Fe}{2}  	 & 1144.938  &   $1.06 \times 10^{-1}$    &	42	& $ 7.3 \pm 2.5 $	& $12.88 \pm 0.15 $	  &  7,8   \\
\enddata
\tablecomments{Uncertainties are $1\sigma$ error. Upper limits indicate that no feature is present and
are $3 \sigma$ estimates, except for \ion{N}{3}. 
Lower limits indicate that the absorption line is saturated.
Colons indicate that the value is uncertain. (See text for more details).\\
($a$) Rest frame vacuum wavelengths and oscillator strengths are from Morton (private communication, 2000).\\
($b$) 1: LiF\,1A MDRS; 2: LiF\,1A LWRS; 3: SiC\,1A LWRS; 4: SiC\,2A LWRS; 5: SiC\,1B MDRS; 
6: SiC\,1B LWRS; 7: LiF\,2A MDRS; 8: LiF\,1B MDRS, 9: SiC\,1A MDRS. Segments
between brackets mean that the lines measured on those different segments 
are consistent with the presented result, but was of lower quality
and are therefore not included in the final results. The S/N level was averaged when more than one 
segment was used (though generally these ratios were similar in the different segments).
($c$) The stellar contribution ($\sim$$30\%$; see text) was removed  from these measurements.\\
($d$) \ion{N}{1} $\lambda$953.970 is blended with \ion{N}{1} $\lambda$954.104 making these values uncertain.\\
($e$) The stellar contribution ($\sim$$7\%$; see text) was removed  from these measurements, but 
\ion{N}{3} $\lambda$989 is also blended with \ion{Si}{2} $\lambda$989.9. This measurement supposes that the line
is not saturated. \\
($f$) This line is a triplet of lines blended at the \fuse\/ resolution. \\
($g$) \ion{Ar}{1} $\lambda$1066.660 is blended with the strong stellar \ion{Si}{4} feature at 1066.6 \AA.
}

\end{deluxetable}
\clearpage

\begin{deluxetable}{lccccc}
\tablecolumns{6}
\tablewidth{0pc} 
\tablecaption{\ion{D}{1} measurements: AOD results \label{t2}} 
\tablehead{\colhead{$\lambda_{\rm lab}$} &   \colhead{$f$} & \colhead{S/N} &\colhead{$W_\lambda$} & \colhead{$\log N_a$}& \colhead{Segment} \\ 
\colhead{\AA} & \colhead{} &\colhead{} &\colhead{(m\AA)}& \colhead{(${\rm cm}^{-2}$)}& \colhead{} }
\startdata
1025.443 &    $7.91 \times 10^{-2}$	&     35      & $ 34.2 \pm 6.8 $	& $ 13.75 \pm 0.09 $	&  LiF\,1A MDRS	\\		 
	 &				&     35      & $ 41.5 \pm 8.7$$^*$	& $ 13.84 \pm 0.10$$^*$	&  LiF\,1A MDRS	\\		 
	 &				&     27      & $ 32.9 \pm 7.0 $	& $ 13.72 \pm 0.10 $	&  LiF\,2B MDRS	\\		 
	 &				&     44      & $ 37.5 \pm 5.8 $	& $ 13.77 \pm 0.08 $	&  LiF\,1A LWRS	\\		 
	 &				&     26      & $ 40.4 \pm 12.7$	& $ 13.80 \pm 0.15 $	&  LiF\,2B LWRS	\\		 
	 &				&     23      & $ 39.3 \pm 9.2 $	& $ 13.81 \pm 0.16 $	&  SiC\,1A LWRS	\\		 
972.272  &    $2.90 \times 10^{-2}$	&     20      & $ 11.5 : (< 14.8) $   & $ 13.67 : (< 13.79)  $	&  SiC\,1B LWRS	\\		 
	 &				&     20      & $ 13.2 : (< 14.8) $   & $ 13.74 : (< 13.79)  $	&  SiC\,1B MDRS	\\		 
	 &				&     23      & $ 18.9 : $	      & $ 13.92 :  $		&  SiC\,2A LWRS	\\		 
949.484  &   $1.39 \times 10^{-2}$	&     27      & $ 10.8 : (< 11.3) $   & $ 13.98 : (< 14.00)  $	&  SiC\,1B LWRS	\\		 
\enddata
\tablecomments{
Uncertainties are 1$\sigma$ error. Upper limits in brackets 
are 3$\sigma$ estimates. Colons indicate that the value is uncertain. (See text for more details).
(*) Stellar continuum, see text for more details.}
\end{deluxetable}

\begin{deluxetable}{lcccc}
\tablecolumns{5}
\tablewidth{0pc} 
\tablecaption{Column densities with $2\sigma$ errors \label{t3}} 
\tablehead{ \colhead{Ions}  & \colhead{AOD}  & \colhead{PF\,1}  & \colhead{PF\,2}  & \colhead{Adopted}}
\startdata
\ion{D}{1}	& $13.79 \pm 0.14$     		& $13.89 \pm\, ^{0.05}_{0.06}$	& $13.86 \pm 0.09$	& $13.85 \pm 0.09$	  \\
\ion{C}{2*}	& $13.06 \pm 0.10$		& \nodata	      		& \nodata		& $13.06 \pm 0.10$	  \\
\ion{N}{1}	& $14.33 \pm\, ^{0.06}_{0.04}$	& $14.31 \pm 0.04$		& $14.37 \pm 0.12$	& $14.34 \pm\, ^{0.09}_{0.08}$	  \\
\ion{O}{1}	& $15.21 \pm\, ^{0.10}_{0.08}$	& $15.29 \pm\, ^{0.06}_{0.05}$	& $15.27 \pm 0.09$	& $15.26 \pm\, ^{0.08}_{0.07}$    \\
\ion{Ar}{1}	& $12.86 \pm 0.14$     		& \nodata			& \nodata		& $12.86 \pm 0.14$	   \\
\ion{Fe}{2}	& $12.88 \pm\, ^{0.26}_{0.38}$	& $13.00 \pm\, ^{0.09}_{0.11}$	& \nodata		& $12.94 \pm\, ^{0.23}_{0.31}$	   \\
\enddata
\tablecomments{AOD gives the results 
of the apparent optical depth method described in \S~\ref{aod}.
PF\,1 and PF\,2 refer to the two independent profile fitting analyses described 
in \S~\ref{pf}.}
\end{deluxetable}

\clearpage

\end{document}